\documentclass{article}

\usepackage[total={6in, 8in}]{geometry}
\usepackage{authblk}
\usepackage{graphicx}
\usepackage{amssymb}
\usepackage{amsmath}
\usepackage{color}
\usepackage{url}
\usepackage{hyperref}

\usepackage[backend=biber, style=numeric, sorting=none]{biblatex}
\DeclareFieldFormat[article]{number}{\addspace#1}

\title{``Layer-by-layer'' unsupervised clustering of statistically relevant fluctuations in noisy time-series data of complex dynamical systems}

\author[1]{Matteo Becchi}
\author[1]{Federico Fantolino}
\author[1,2]{Giovanni M. Pavan\thanks{To whom correspondence should be addressed. E-mail: giovanni.pavan@polito.it}}

\affil[1]{Department of Applied Science and Technology, Politecnico di Torino, Torino 10129, Italy}
\affil[2]{Department of Innovative Technologies, University of Applied Sciences and Arts of Southern Switzerland, Lugano-Viganello 6962, Switzerland}

\date{\today}

\begin{document}

\maketitle

\begin{abstract}
Complex systems are typically characterized by intricate internal dynamics that are often hard to elucidate. Ideally, this requires methods that allow to detect and classify in unsupervised way the microscopic dynamical events occurring in the system. However, decoupling statistically relevant fluctuations from the internal noise remains most often non-trivial. Here we describe ``{\em Onion Clustering}'': a simple, iterative unsupervised clustering method that efficiently detects and classifies statistically relevant fluctuations in noisy time-series data. We demonstrate its efficiency by analyzing simulation and experimental trajectories of various systems with complex internal dynamics, ranging from the atomic- to the microscopic-scale, in- and out-of-equilibrium. The method is based on an iterative {\em detect-classify-archive} approach. In similar way as peeling the external (evident) layer of an onion reveals the internal hidden ones, the method performs a first detection and classification of the most populated dynamical environment in the system and of its characteristic noise. The signal of such dynamical cluster is then removed from the time-series data and the remaining part, cleared-out from its noise, is analyzed again. At every iteration, the detection of hidden dynamical sub-domains is facilitated by an increasing (and adaptive) relevance-to-noise ratio. The process iterates until no new dynamical domains can be uncovered, revealing, as an output, the number of clusters that can be effectively distinguished/classified in statistically robust way as a function of the time-resolution of the analysis. {\em Onion Clustering} is general and benefits from clear-cut physical interpretability. We expect that it will help analyzing a variety of complex dynamical systems and time-series data. 
\end{abstract}

\newpage

\section*{Introduction}
Understanding the dynamics of complex systems is typically a hard task and presents inherent challenges. Cause-and-effect relationships, as well as the spatial and temporal correlations, are often hidden within the noise generated by a large number of units that dynamically communicate with each other in an intricate network of interactions~\cite{sattari2022modes, liu2021quantifying, borge2016dynamics, nitzan2017revealing, basak2021information, crippa2022molecular}. The behavior of these systems is often controlled by local (rare) fluctuations, but detecting and distinguishing them from the intrinsic noise of datasets extracted from their trajectories is often non-trivial~\cite{hong2008unsupervised}. This holds for a variety of systems across different scales, from the atomic- and molecular- to the macroscopic-level~\cite{cho2021dynamics}. 
The relevance of local microscopic fluctuations has been shown, for example, in studies of metal surfaces and nanoparticles~\cite{baletto2019structural, cioni2023innate}, supramolecular fibers~\cite{gasparotto2019identifying, bochicchio2019defects, de2021controlling}, and nucleation processes~\cite{wolde1997enhancement, lutsko2019crystals}. On a macroscopic scale, the effects of local fluctuations and events on the behavior of the whole system are seen in collective phenomena such as, {\it e.g.}, bird flocks~\cite{nagy2010hierarchical, cavagna2010scale, attanasi2014information}, fish banks~\cite{butail2016model, porfiri2018inferring}, as well as in the dynamics of economic and stock market systems~\cite{mantegna1999introduction, duan2022network, liu2021quantifying}. 
The study of the behavior of these complex systems over time, by either computer simulations or experimental setups, typically generates a large amount of multivariate data that are often non-trivial to analyze. In particular, extracting meaningful and interpretable information from such noisy time-series is generally hard. 
To address this issue, common strategies involve the use of either knowledge-based or data driven descriptors. Such descriptors serve as a crucial intermediary step, effectively reducing the amount of data to an interpretable form and facilitating the extraction of useful information for elucidating the underlying dynamics.

Structural descriptors -- either specific and knowledge-based, or abstract, general ones -- are often used to extract comprehensive insights into the structural features of complex systems. As an example, general high-dimensional structural descriptors, such as, {\it e.g.}, the Smooth Overlap of Atomic Positions (SOAP) descriptor~\cite{bartok2013representing}, have been recently used to obtain a data-driven structural characterization of, {\it e.g.}, water and ice systems~\cite{monserrat2020liquid, donkor2023machine, offei2022high, ansari2020insights, capelli2022ephemeral}, metallic~\cite{cioni2023innate, crippa2023machine}, ionic~\cite{lionello2022supramolecular}, and soft (biological or artificial) molecular systems~\cite{gasparotto2019identifying, capelli2021data, gardin2022classifying, cardellini2023unsupervised} from molecular dynamics (MD) simulations. 
However, at the same time, pattern recognition analyses based on such structural descriptors typically struggle in capturing infrequent dynamical events and local fluctuations that play a pivotal role in determining their behavior~\cite{sharp2018machine, cho2021dynamics, cioni2023innate, caruso2023timesoap, crippa2023detecting}. Conversely, it has been demonstrated how time-series analyses tracking the temporal evolution and fluctuations of descriptors in time allow retaining a richer amount of information all the events occurring in complex molecular systems. 
One recent example is the time-SOAP ($t$SOAP) descriptor, which measures the rate of change of the SOAP power spectrum of each unit in a multi-unit trajectory of a dynamical molecular system~\cite{caruso2023timesoap}. A time-series analysis of $t$SOAP was recently shown to retain rich information information of the structural change events that occur within molecular systems, including rare local events. Another example is the Local Environments and Neighbor Shuffling (LENS) descriptor, which tracks changes in the identity of the neighbor units that surround every unit in a dynamical network~\cite{crippa2023detecting}. While these examples show the potential of studying the behavior of complex systems based on the trajectories of their individual units over time, this shifts the focus from pattern recognition on global datasets to the study of time-series data and of their dynamical fluctuations. 

One key challenge in time-series analysis is clustering~\cite{rai2010survey, rodriguez2014clustering, pizzagalli2019trainable, barrio2023clustering}, and in particular the identification and classification of fluctuations that are relevant against the background noise~\cite{keogh2002finding, gupta2013outlier}. Unsupervised clustering algorithms frequently struggle in identifying rare events and sparse fluctuations due to their negligible statistical weight, and because the detection of more populated clusters implicitly sets a metric that is too coarse to discriminate well less populated ones. Typically, the higher the density of certain clusters, the more difficult is to detect and classify the less populated ones. Detecting and retaining information on relevant fluctuations, separating them from noise, is of key importance to reconstruct the physics of the studied systems~\cite{fernex2021cluster}. 

Furthermore, this is particularly relevant in complex systems, whose collective and adaptive behaviors often emerge locally (both in time and space) and are intimately related to rare events and local fluctuations~\cite{bochicchio2019defects, albertazzi2014probing}. Unsupervised approaches capable of providing a microscopic analysis of time-series {\it via} systematic and robust detection and clustering of the fluctuations occurring within them would be desirable to this end. 
However, the most common clustering algorithms are either built to handle static datasets, or to perform whole time-series clustering~\cite{wang2006characteristic, langkvist2014review, madiraju2018deep, javed2023somtimes}, and are thus not well-suited to obtain a single-point (microscopic-level) clustering of the local dynamical events occurring in the time-series~\cite{aghabozorgi2015time, keogh2005clustering, bogetti2023lpath}. 

Here we introduce {\em Onion Clustering}, a general, simple, unsupervised, and physically interpretable algorithm tailored for single-point clustering of fluctuations in noisy time-series data. Our approach is founded on the general concept that every (microscopic) environment in a system is characterized by an average dynamics and by a characteristic noise (amplitude of fluctuations around the mean). 
As a core idea, the algorithm is based on an iterative {\em detect-classify-archive} approach where, step-by-step, the highest-density microscopic dynamical environment present in the system is detected, its dynamical features (average dynamics and characteristic noise) are classified, and its signal (and the related noise) is then removed from the time-series, which is then analyzed again according to the same iterative procedure. In similar way as peeling the external layer of an onion reveals the internal hidden ones, after the classification of the evident dynamical environments, at every iteration the method can efficiently uncover and classify the hidden (least populated) dynamical domains thanks to an iteratively enhancing signal-to-noise ratio. In this way, {\em Onion Clustering} allows extracting all the features that can be effectively classified in a time-series. Noteworthy, instead of leaving the user to make an {\it a priori} choice on the resolution to be used for an analysis (which is critical, and typically requires a prior knowledge of the system under study), {\em Onion Clustering} reveals as an outputs the number of clusters that can be effectively classified in a statistical robust way in a time-series as a function of the time-resolution used in the analysis. This provides a robust unsupervised clustering algorithm with a non-common physical interpretability that allows for a transparent intuition into the mechanism of clusters detection and an informed interpretation of the obtained results. 

We demonstrate the efficiency and generality of {\em Onion Clustering} by analysing a variety of complex dynamical systems, ranging from the microscopic to the mesoscopic scales, with diverse internal dynamics, in- and out-of-equilibrium conditions. {\em Onion Clustering} is open-source~\cite{onion-git, gmp-git}, and is released as a Python3 package~\cite{onion-pypi}. We expect that this method will constitute a precious tool to study complex dynamical systems in general, and the microscopic events occurring within them and controlling their behavior. 

\section*{Results and discussion}
\subsection*{The method and a test on water-ice dynamic coexistence}

\begin{figure*}[!ht]
    \centering
    \includegraphics[width=\textwidth]{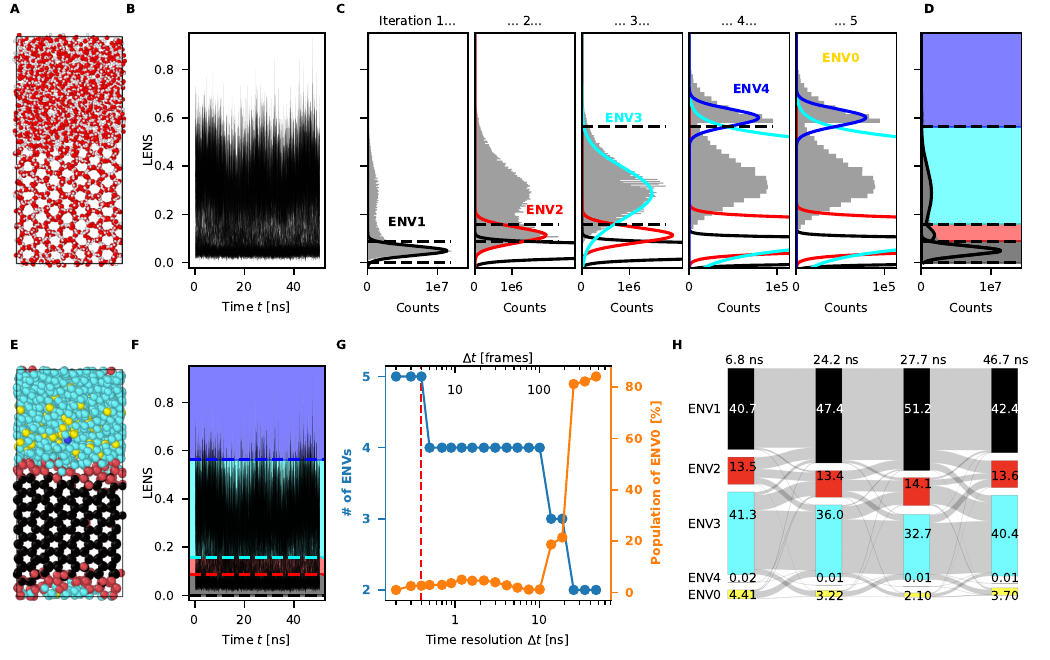}
    \caption{\textbf{Clustering of LENS signals on ice/water coexistence MD simulation}. A: Snapshot of the simulation of ice/water coexistence; the simulation is performed on 2048 TIP4P/ICE molecules, lasts $50$~ns and it's sampled every $0.1$~ns. B: LENS signals for all the oxygen atoms, as a function of time. C: Data cumulative histograms at the five iterations of the algorithm, using a time resolution $\Delta t = 0.4$~ns. The solid lines are the Gaussian best fit of the maximum of the distribution. The dashed lines are the threshold between the identified clusters. At the fifth iteration, no data are assigned to the proposed cluster, and the algorithm stops. D: final clustering of the LENS signals. E: Snapshot of the simulation, colored according to the clustering. F: Same LENS signals of panel B; background is colored according to the thresholds given by the clustering algorithm. G: Sankey diagram between four different times along the simulation. Colored bars are proportional to the clusters' populations, gray lines are proportional to the number of molecules moving from one cluster to another. H: Blue line: number of clusters identified as a function of the time resolution $\Delta t$; orange line: percentage of the data points classified in the ENV0 cluster as a function of the time resolution $\Delta t$. The red dashed line at $\Delta t = 0.4$~ns indicates at which time resolution the analysis shown in the previous panels was performed. }
    \label{fig:fig1}
\end{figure*}

In this section, we illustrate the algorithm using as a first demonstrative case the clustering of data extracted from a $50$~ns long MD simulation trajectory containing 2048 TIP4P/ICE~\cite{abascal2005potential} molecules ($1:1$ liquid water:ice) in dynamic equilibrium in correspondence of the melting temperature. A complete description of the algorithm is provided in the Methods section and in the Supplementary Information (SI). 

Fig~\ref{fig:fig1}A shows, as an example, the LENS signals~\cite{crippa2023detecting} for all 2048 individual water molecules in the system sampled every $\Delta t=0.1$~ns. 
The input dataset in this example thus consists of $N=2048$ univariate time-series $x_i(t)$, $1\leq i\leq  N$ labelling the water molecules in the simulation trajectory, each containing $0\leq \Delta t < T$ sampled time-steps. We underline that the same analysis can be conducted also using different descriptors -- see, {\it e.g.}, Fig~S1 for the consistent results obtained using the $t$SOAP descriptor~\cite{caruso2023timesoap}: as it is shown in the next sections, {\em Onion Clustering} is general and can be applied virtually to any time-series data. 

LENS is a permutation-invariant descriptor that measures how much the neighborhood of each (water) molecule in the system changes in term of neighbor molecular individuals in the sampling time-interval ($\Delta t=0.1$~ns in this case). In detail, LENS captures local events such as reshuffling, addition, or loss of neighbor molecules, and it can be thought of as a local dynamicity parameter related, in some sense, to a local diffusivity of the molecules in the system. 
The LENS signal is thus expected to be lower in the solid phase (ice) and higher in the liquid water phase, where the molecular rearrangement is faster. 
Fig~\ref{fig:fig1}B shows the LENS signals time-series. It is worth mentioning that typical unsupervised pattern recognition approaches used to analyze the entire dataset basically detects two main environments -- liquid water and solid ice -- in such a system~\cite{monserrat2020liquid, zeni2022exploring, capelli2022ephemeral, caruso2023timesoap}, where both states are well represented statistically (see also leftmost panel of Fig~\ref{fig:fig1}C: two peaks in the density of the signals at LENS values of~$\sim0.05$ and $\sim0.3$)~\cite{crippa2023detecting}. 
However, this becomes problematic in cases where there are states/environments that are present in a low fraction and that are typically overlooked in pattern recognition analyses due to their negligible statistical weight.
Similarly, for the same reasons such approaches struggle in providing information on the (rare) transitions between the main states and on the involved intermediate transition states.
On the other hand, recently it has been demonstrated that studying the time-series of such signals allow detecting and retaining information also of rare/local transition events that appear as outliers in the time-series~\cite{crippa2023detecting, caruso2023timesoap}. However, to what extent one fluctuation is different from noise or from another fluctuation, how similar/different the various fluctuations are, and, in particular, with what statistical confidence it is thus possible to group them based on their similarity are typically non-trivial questions.

Using this as a first demonstrative case, we show how {\em Onion Clustering} is capable of performing a microscopic analysis of the time-series, subdividing them into different dynamical environments whose fluctuations have characteristic fingerprints in terms of intensities and oscillation amplitudes. The algorithm automatically identifies in an unsupervised way the dynamical micro-clusters that may be present in the system (the number of clusters is thus not set {\it a priori}, but is rather an output of the algorithm) and assign points to them, assessing their difference/similarity in a statistically robust way. 
The method follows an ``onion-like'' approach, where the environments that are more evident/certain are first detected and classified and, after removing them from the signal, the algorithm proceeds iteratively in classifying the less-evident/hidden ones. 
In particular, in a first iteration (``Iteration 1''), {\em Onion Clustering} starts by computing the cumulative histogram of all the data points in the time-series (leftmost panel of Fig~\ref{fig:fig1}C). The global maximum of the histogram is then identified: in this test case, the LENS signals have the maximum density at LENS~$\sim0.05$ (a relatively low value, corresponding to the solid-ice phase: {\it vide infra}). 
The idea behind the algorithm is to assume that each maximum of the histogram corresponds to one well determined dynamic environment in the system, which is thus characterized by an average dynamics (average LENS value) and by a normally-distributed characteristic noise. 
Based on this concept, once identified the highest density peak, the algorithm fits a Gaussian distribution of the form 
\begin{equation}
    \label{eq:eq1}
    P(x) = \frac{A}{\sqrt{\pi}\sigma}\exp{\left[-\left(\frac{x - \mu}{\sigma}\right)^2\right]}
\end{equation}
to the histogram maximum, as shown in Fig~\ref{fig:fig1}C (black solid curve). The mean $\mu$, the (rescaled) standard deviation $\sigma$ and the area $A$ of the Gaussian are the fit parameters. 

This identifies a first dynamical environment, labelled as ``ENV1'', which is characterized by LENS values within the interval $\left[\mu - 2\sigma, \mu + 2\sigma\right]$ (and that in this case identifies the solid ice domain). This criterion is equivalent to discard data points that do not belong to ENV1 with a probability higher that $99.5\%$.
As a next step, the algorithm slices the time-series in consecutive (non-overlapping) time-windows of length $\Delta t$. 
The algorithm identifies all the molecules that remain always in ENV1 (without jumping in/out ENV1) in $\Delta t$, for all $\Delta t$s along the trajectory, thus classifying all molecules that, at the resolution of the analysis ($\Delta t$), appear as persistently belonging to ENV1 for time intervals at least equal to (or multiple of) $\Delta t$. 
After this step, all these already classified ENV1 signals are removed from the data and the time-series is analyzed again in another iteration. 

It is worth noting that $\Delta t$ is the only parameter required by {\em Onion Clustering}. In time-series analysis the choice of the $\Delta t$ is critical, as it sets {\it de facto} the time-resolution in the analysis (as it will be discussed in more detail below). Larger values of $\Delta t$ correspond to a lower resolution, while smaller values of $\Delta t$ correspond to a higher resolution in the study of the time-series, respectively reducing/enhancing the discretization of the events that occur along the studied trajectory. To prevent the use of the algorithm by the users as a black box (or leveraging too much prior assumptions/knowledge on/of the system), {\em Onion Clustering} performs the analysis at many different $\Delta t$s and outputs the results that can be effectively obtained at the different resolutions (see next section for a detailed discussion on the effect of changing the $\Delta t$ in the analysis). 
As a demonstrative case, here in Fig~\ref{fig:fig1} we show the results obtained by using a $\Delta t = 0.4$~ns, which corresponds to 4 simulation time-frames in the analysis of water-ice molecules that coexist in dynamic equilibrium (results obtained with other $\Delta t$ values are available in Fig~S2 in the SI). 

The second iteration starts again by computing the cumulative histogram of the data points. As can be seen in the ``Iteration 2'' panel of Fig~\ref{fig:fig1}C, the removal of the points classified into ENV1 changes the histogram. Now, environments that before were difficult to identify as hidden by the ENV1 data/noise become the new prominent features of the histogram. In this way, by identifying and removing a new environment at each iteration, the algorithm automatically adjusts the data range in order to improve its efficiency in identifying environments which are less and less statistically relevant (note, in fact, the finer and finer scale on the $x-$axes of Fig~\ref{fig:fig1}C). 

The algorithm then proceeds exactly as in the previous iteration. The global maximum is identified, and fitted with a Gaussian distribution (solid red line in Fig~\ref{fig:fig1}C), which gives the limit of the new environment, ``ENV2''. Then, the data-windows entirely included into ENV2 are detected, stored and removed. The same procedures is repeated iteratively. As it is shown in Fig~\ref{fig:fig1}C, in this specific system at this resolution ($\Delta t = 0.4$~ns) four environments can be identified, which are characterized by increasing values (and lower densities) of LENS signal. 

Such {\it find-classify-archive} strategy builds on a hierarchical certainty approach that classifies first the data that are more certain and then, layer-by-layer, proceeds in classifying hierarchically the remaining part of the time-series. Noteworthy, eliminating the ENV1 data after the classification results also in the deletion of the associated noise, which augments in the next iteration the sensitivity of the method and the relevance-to-noise ratio. At the fifth iteration, a new environment ``ENV5'' is fitted. But no signal window in the remaining dataset is entirely included within it: {\it i.e.}, there are no molecules that stay into such ENV5 at least for the duration of $\Delta t = 0.4$~ns. The algorithm thus meets a termination condition, and the iterative process stops. The remaining data points, which were not classified into any of the previously identified environments (at least with this choice of $\Delta t$) are classified as a last cluster, labelled as ``ENV0''. ENV0 contains all the data that are not persistently part of ENVs1-4 for at least $\Delta t$ ({\it e.g.}, transitions). 
The key importance of such ENV0 environment from the physical, statistical, and methodological points of view is discussed in detail in the next section. 

Once the iterative analysis terminates, the algorithm determines the thresholds between the different environments, defined as the intersection points between the various Gaussian distributions (Fig~\ref{fig:fig1}D-F: dashed lines).
This identifies the main ENV1-4 clusters colored in Fig~\ref{fig:fig1}D-F. 
In this specific case, the algorithm finds 4 statistically relevant LENS environments (ENVs1-4), along with the ENV0 cluster. The characterization of the LENS signals within each cluster is displayed in Fig~S3 in the SI. As can be seen from the simulation snapshot in Fig~\ref{fig:fig1}E and in Supplementary Movie~S1, the cluster ENV1 corresponds to the solid ice phase, ENV2 to the solid/liquid interface (ice surface), ENV3 comprises the majority of the molecules in the liquid water phase, while ENV4 contains a smaller fraction of water molecules that, as described recently~\cite{capelli2022ephemeral}, may occasionally form ephemeral ice-like clusters that in such conditions continuously freeze and re-melt in the liquid domain. 

\subsubsection*{The key importance of time-resolution}
Changing the time resolution of the analysis, $\Delta t$, determines what type of information can be effectively captured and how much information is lost. Setting the $\Delta t$ means setting the sensitivity and uncertainty in the analysis, in that the resolution is sufficient to classify some events but not other (faster) ones. This reflects in the number of clusters (ENVs) that are classified by the analysis. For example, reducing the $\Delta t$ increases the resolution in the study of the time-series, and results in an augmented discretization and a higher number of detected clusters (ENVs). At the same time, the amount of information that remains ``undetermined'' at a given $\Delta t$ is also exactly quantified by the ENV0 cluster. In particular, the higher is the data content of the ENV0 cluster, the higher is the amount of information that cannot be classified in a statistically robust way. 
In this specific case, where $\Delta t=0.4$ ns, the molecules belonging to the ENV0 cluster and corresponding to fast transitions between the ENVs1-4 environments weight $\sim 3.5\%$ of the total data points. 

As anticipated above, instead of making an {\it a priori} choice of the time-resolution -- typically leveraging on a considerable prior knowledge of the system by an expert user, or on a ``blind'' unsupervised choice that risks to make the software a ``black box'' -- {\em Onion Clustering} uses a different strategy that improves its transparency and physical interpretability. 
In particular, the software always performs the analysis at different values of $\Delta t$, ranging from the maximum resolution of $\Delta t = 2$~frames, to the minimum one, corresponding to $\Delta t = T$, where $T$ is the entire time-series (the latter case results in a typical pattern-recognition analysis conducted on the entire dataset). In this demonstrative case, the analysis is conducted ranging from $\Delta t = 0.2$~ns ($2$~frames) to $\Delta t = 47$~ns ($470$~frames, comparable with the entire trajectory length). 
At every usage, {\em Onion Clustering} outputs a plot such that of Fig~\ref{fig:fig1}G. In blue and orange are respectively shown the number $n$ of statistically relevant clusters that can be classified in a robust way (ENV1-to-n) and the fraction (in \%) of unclassified data contained in the ENV0 cluster as a function of the $\Delta t$. 
For smaller $\Delta t$ values (up to $\Delta t = 0.4$~ns) 5 clusters are found (4 statistically relevant ones -- ENV1-to-4 -- plus the ENV0, which collects the unclassified data points). For intermediate $\Delta t$ values ($0.5 \leq \Delta t \leq 10$~ns), the ENV clusters reduce to 4. Reducing the resolution of the analysis (increasing the $\Delta t$), ENV4, which corresponds to molecules with very high LENS values (identifying ephemeral ice-like domains forming/dissolving in the liquid water), merges with ENV3 (corresponding to liquid water: see also Fig~S2 in the SI). Evidently, the resolution is no more high enough to discriminate such molecules from liquid ones. Noteworthy, this outcome is also physically relevant, because it reveals the maximum time-scale at which such ephemeral ice-like domains can be effectively discriminated from a statistical point of view and provide a rough information on their survival lifetime (which is shorter than $500$~ps). 

Increasing further the $\Delta t$ ($>10$~ns) starts producing a loss of information that can be effectively classified. It is not possible anymore to distinguish ENV2 -- {\it i.e.}, the solid/liquid interface -- as a distinct cluster, and only ENV1 (solid ice) and ENV3 (liquid water), along with the unclassified ENV0 cluster, can be identified. This outcome provides a qualitative estimate for the average lifetime of a water molecule in ENV2 ($0.5-10$~ns), which is compatible with previous studies on the diffusion coefficient of water molecules at the ice/water interface~\cite{karim1988ice}. 

It is worth noting how for $\Delta t > 10$~ns the fraction of data in the ENV0 cluster (unclassified data) sharply increases. This indicates that the time resolution starts to be insufficient to reconstruct the microscopic physics of the system, and a significant fraction of data points remain unclassified during the iterative process. 
In particular, for $\Delta t > 20$~ns the sole environments that can be detected are ENV1, corresponding to the bulk of solid ice (molecules that do not diffuse along the entire simulation) and ENV0, gathering in this case $\sim 80\%$ of the total data points, which includes all molecules that move in the system. 
Interestingly, in this case the result of the analysis becomes consistent with the typical result that is obtained via unsupervised clustering approaches on datasets extracted from the entire trajectory~\cite{monserrat2020liquid, zeni2022exploring, capelli2022ephemeral, caruso2023timesoap}. 

The plot of Fig~\ref{fig:fig1}G is a key feature of {\em Onion Clustering}, providing relevant information. On the one hand, it sheds light on the physics underlying the system under study. On the other hand, it provides important information on the performance of the clustering algorithm and on the robustness of the classification that this provides. The correlation between the number of detected clusters and $\Delta t$ unveils the characteristic time-scales of the various dynamic environments and the transitions occurring within the system. Conversely, the correlation between the population of the ENV0 cluster and $\Delta t$ indicates the time resolution at which the algorithm begins to struggle, having insufficient resolution and statistics to classify large parts of the time-series.
The plot of Fig~\ref{fig:fig1}G is a key output of {\em Onion Clustering} in that it provides the user with a statistically-robust litmus paper useful to choose {\it a posteriori} the resolution of the analysis depending on the type of events that one wants to study (instead of {\it a priori}, {\it e.g.}, based on human-based assumptions). 
This is a non-common feature for a fully unsupervised method, which in this way gets rid of ``black box'' issues/limitations and gains physical interpretability.
Instead of attempting to fit all data into clusters, the philosophy of {\em Onion Clustering} is to determine the amount of information that cannot be statistically classified at a given resolution (starting from the concept that every measurement method has intrinsic limitations that cannot be neglected), to subtract it, and to classify only the data that can be effectively classified from a statistical point of view. This is key, as it provides an advantage in terms of transparency, reliability, robustness, and repeatability of the analysis. 

\subsubsection*{Characterizing the microscopic dynamics of the system}
Having assigned every data point to one of the identified clusters, it is easy, {\it e.g.}, to track not only how the different clusters populations vary with time, but also the transitions of the individual water molecules between the various environments along the time-series. In the Sankey diagram of Fig~\ref{fig:fig1}H, the height of the colored bars is proportional to the populations of the 5 detected clusters at 4 different representative time steps taken along the trajectory. The gray bands between the time steps provide a coarse-grained representation of the number of molecules that moved between any pair of clusters in the time-interval between the two represented snapshots. While the diagram of Fig~\ref{fig:fig1}G is here purely demonstrative, and it shows just the departure and arrival clusters for water molecules between distant time intervals, a more detailed characterization of the exchange pathways and of the inner microscopic dynamics of the system can be easily attained by tracking the transitions between shorter time intervals. 
Nonetheless, this plot clearly shows that, as expected, the exchange of water molecules between solid and liquid phases occurs mainly {\it via} an intermediate dynamical environment ({\it i.e.}, via the ice/water interface). The unclassified (ENV0) events occur mainly in connection with the liquid phase. This indicates that, among the various transitions that this encompasses, considerable part of ENV0 is related to local ephemeral ice-like domains that fastly form/dissolve in the liquid domain in these conditions~\cite{capelli2022ephemeral} (events that occurs too fast to be statistically classified as a distinct cluster at the time resolution of $\Delta t=0.4$ ns). 

\subsection*{Different test cases in different conditions}
The results discussed above refer to a case of a system in dynamical equilibrium, with a rather ``fluid'' internal dynamics and characterized by exchange events taking place between similarly-populated liquid and solid phases. While this is a particular case, to prove the generality of the method we tested {\em Onion Clustering} on time-series obtained from a variety of systems with diverse internal dynamics: {\it e.g.}, systems far from the equilibrium, or dominated by local rare fluctuations. Finally, to prove the broad applicability of the algorithm, this is also tested on multivariate/multidimensional time-series extracted both from synthetic and experimental datasets. 

\subsubsection*{{\em Onion Clustering} of out-of-equilibrium time-series: water freezing}

\begin{figure*}[!ht]
    \centering
    \includegraphics[width=\textwidth]{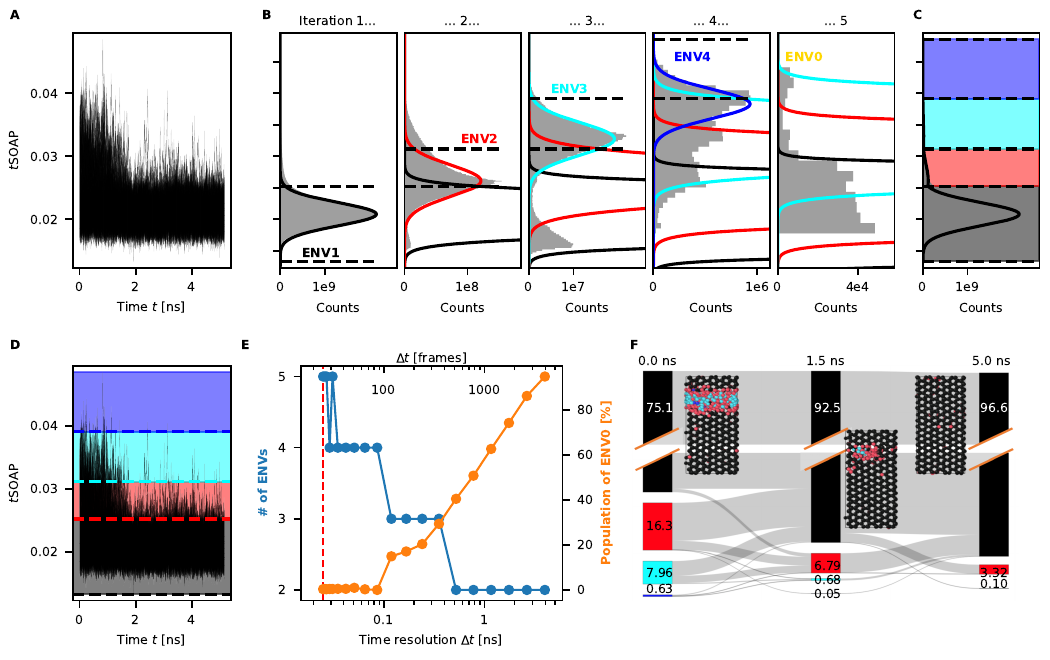}
    \caption{\textbf{Analysis of out-of-equilibrium time-series: freezing water.} A: $t$SOAP values for 2048 TIP4P/ICE molecules, along the $5$~ns long MD simulation sampled every ps, smoothed with a moving average with width $25$~frames ($25$~ps). B: Data cumulative histograms at the five iterations of the algorithm. The solid lines are the Gaussian best fit of the maximum of the distribution. The dashed lines are the thresholds between the identified clusters. At the fifth iteration, the Gaussian fit does not converge, and the algorithm stops. C: final clustering of the $t$SOAP signals. D: The $t$SOAP signals; background is colored according to the threshold given by the clustering. E: Blue line: number of clusters identified as a function of the time resolution $\Delta t$; orange line: percentage of the data points classified in the ENV0 cluster as a function of the time resolution $\Delta t$. The red dashed line at $\Delta t = 0.025$~ns indicates at which time resolution the analysis shown in the previous panels was performed. F: Sankey diagram between three different times along the simulation. Colored bars are proportional to the clusters' populations, gray lines are proportional to the number of molecules moving from one cluster to another. }
    \label{fig:fig2}
\end{figure*}

Analyzing time-series data that are out-of-equilibrium poses additional challenges compared to well-sampled equilibrium trajectories. Short-lived clusters and transient states may rapidly emerge and disappear, representing only a small fraction of the data with a negligible weight on the entire dataset. Furthermore, in such systems the result of pattern recognition approaches conducted on the entire trajectory changes over time (as the density of the states does). In fact, information on transient states that, {\it e.g.}, may emerge only in the beginning of the time-series disappearing rapidly, but that may be key to understand the evolution of the system, is lost when the time-series that is analyzed becomes longer and longer.
Retaining information on these short-lived states is essential for understanding the time evolution of a system, but keeping memory of these, or in some cases even realizing that they ever appeared during a trajectory, is not always easy. 

We thus tested {\em Onion Clustering} in a prototypical test case where, starting from the equilibrium condition of Fig~\ref{fig:fig1}, the temperature in the system is reduced to $T=267$ K. Such temperature is below the melting point of the TIP4P/ICE model (see Methods section for details). 
In this case, the simulation trajectory that is analyzed is approximately $5$~ns long, with sampling every $1$~ps (for a total of $\sim5000$~frames), which is a sufficient time to observe the entire system freezing. 
As an example, we computed the $t$SOAP descriptor~\cite{caruso2023timesoap} for all water molecules in these out-of-equilibrium trajectories (for comparison with the same system at the equilibrium, see Fig.~S1). In brief, $t$SOAP is a scalar descriptor that measures the rate of variation of the SOAP spectrum~\cite{bartok2013representing} of all molecules in the system. It thus gauges the rate of structural rearrangement within the atomic environment of the molecules: lower in the ice phase, and higher in liquid water. The resulting time-series are shown in Fig~\ref{fig:fig2}A. In this plot, it can be seen that the highest values of $t$SOAP ($> 0.03$), identifying molecules in the liquid phase (faster structural rearrangement of their neighbors), tend to disappear after $\sim2$~ns, leaving only the lower $t$SOAP values corresponding to molecules in the solid ice phase. 
As can be seen in the leftmost cumulative histogram of Fig~\ref{fig:fig2}B, already after $\sim5$~ns the statistical weight of the data points with $t$SOAP~$>0.03$ is negligible, which makes it hard to detect, in analysis conducted on the entire dataset, that there has even been liquid water in this system (and the problem becomes more severe if the simulation last longer). 

Fig~\ref{fig:fig2}B shows the iterations of {\em Onion Clustering} (here as an example, the results obtained using a $\Delta t = 25$~ps are shown). The classified clusters are shown in Fig~\ref{fig:fig2}C-D. Also in this case, at most five environments can be identified in the time-series, corresponding respectively to the ice, the ice/water interface, and two liquid water micro-environments with different $t$SOAP values (and that can be discriminated only at high resolution), along with the ENV0 cluster encompassing the unclassified data points. The significance of these clusters becomes evident in the simulation snapshots shown in Fig~\ref{fig:fig2}F and in the Supplementary Movie~S2. Noteworthy, using $\Delta t = 25$~ps the algorithm accurately classifies liquid water and the ice/water interface, despite these environments vanishing after only $2$~ns of simulation. 

Fig~\ref{fig:fig2}E shows the number of clusters and the population of the ENV0 cluster as a function of the $\Delta t$. The number of clusters decreases from a maximum of 5 for $\Delta t < 40$~ps to $2$ for $\Delta t > 0.5$~ns. At the same time, the fraction of unclassified data in the ENV0 cluster remains negligible up to $\Delta t = 0.1$~ns, while it begins to rise increasing the $\Delta t$, since the time-resolution is insufficient to track the fast evolution of the the system. Notably, this $\Delta t$ value is considerably lower than that observed in the previous section (see Fig~\ref{fig:fig1} for the LENS analysis and Fig~S1 for the $t$SOAP analyses of the equilibrium system). Such a discrepancy is essentially due to a different relevance/noise ratio between the $t$SOAP and LENS descriptors and to the fact that the events become faster when the system evolves rapidly far-from-the-equilibrium. Anyways, such a test shows how also in this case {\em Onion Clustering} reveals in automatic and unsupervised way the resolution necessary to statistically characterize the events that occurred in the beginning of the trajectory, not only providing information that are non trivial to retain but also a physical anchor to prove their relevance and robustness. 

In this test case, an examination of the cluster populations and exchange rates in Fig~\ref{fig:fig2}F offers a deeper insights, clearly demonstrating the out-of-equilibrium behavior observed in the trajectory. As the simulation time progresses, the proportion of liquid water diminishes, then followed by the interface and ultimately leading to their disappearance, while the majority of molecules undergo transition to the solid phase. 

\subsubsection*{Rare local events in time-series: atomic dynamics on metal surfaces}

\begin{figure*}[!ht]
    \centering
    \includegraphics[width=\textwidth]{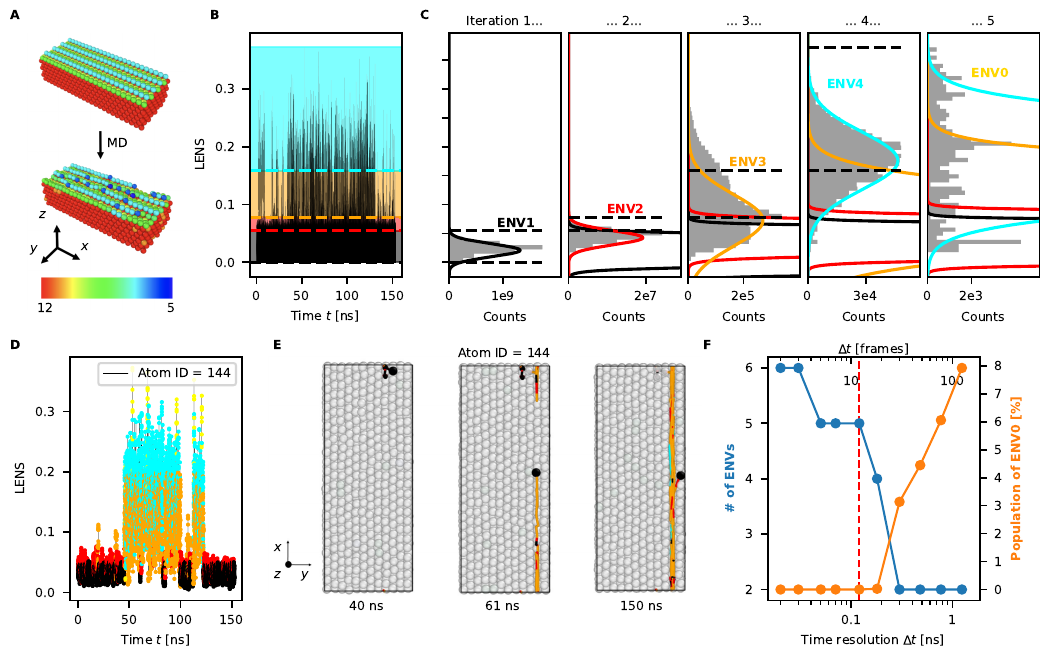}
    \caption{\textbf{Analysis of time-series dominated by rare events.} A: Snapshots of the MD simulation of Cu surface composed by 2400 atoms at $T=600$~K; atoms are colored according to their coordination number. The upper snapshot is at $T = 0$ $K$, the lower one during the simulation at $T = 600$ $K$. B: LENS values for all the Cu atoms, along the $150$~ns long simulation sampled every $10$~ps, smoothed with a moving average with width 10 frames. C: Data cumulative histograms at the five iterations of the algorithm. The solid lines are the Gaussian best fit of the maximum of the distribution. The dashed lines are the threshold between the identified clusters. After the fifth iteration, no data are assigned to the proposed cluster, and the algorithm stops. D: the LENS signal for the atom with ID = 144, colored according to the cluster it belongs at each frame. E: Top view of the simulation box, at three different times $t=40, 61$ and $150$~ns. The atom with ID = 144 is highlighted in black, and its trajectory up to that point is colored according to its environment. F: Blue line: number of clusters identified as a function of the time resolution $\Delta t$; orange line: percentage of the data points classified in the ENV0 cluster as a function of the time resolution $\Delta t$. The red dashed line at $\Delta t = 0.12$~ns indicates at which time resolution the analysis shown in the previous panels was performed. }
    \label{fig:fig3}
\end{figure*}

Another scenario where clustering algorithms often struggle is in detecting amidst background noise and classifying rare events and local fluctuations that may be dominant but have a negligible statistical weight. As a prototypical example of such a system, we tested {\em Onion Clustering} on LENS time-series extracted from an atomistic MD simulation trajectory of a FCC(211) copper surface consisting of $2400$ Cu atoms. The simulation is conducted at a temperature $T = 600$~K using a deep-potential neural network force field that has been recently reported~\cite{cioni2023innate} (see Methods section for details). The MD trajectory lasts $150$~ns and is sampled every $10$~ps (for a total of $15000$~frames). As shown in Fig~\ref{fig:fig3}A, it is known that in this system, while the majority of the surface atoms vibrate within the atomic lattice, a small number of sparse atoms may undergo rapid long-distance sliding motion on the Cu surface~\cite{cioni2023innate,crippa2023detecting,crippa2023machine}. Specifically, such sliding motions are well captured by the LENS descriptor, which has been computed for all atoms along the trajectory, obtaining the time-series of Fig~\ref{fig:fig3}B (LENS values $\gtrsim 0.1$ identify atomic sliding events). 

We performed an {\em Onion Clustering} on these LENS time-series (Fig~\ref{fig:fig3}B-C), using a time resolution for the analysis of $\Delta t = 0.12$~ns (equal to 12 simulation frames). Four statistically relevant LENS environments are identified (ENV1-4), along with the ENV0 cluster. Fig~\ref{fig:fig3}B shows the thresholds between the LENS environments/clusters. 
ENV1 and ENV2 together encompass $\sim 99.95\%$ of the data points, which correspond to static bulk and surface atoms in the system. Remarkably, despite this issue, the algorithm is able to correctly assign the remaining data points to the other microscopic dynamical environments (ENV3-4), which identify the rapid sliding motion of some atoms on the Cu surface. 
Fig~\ref{fig:fig3}D shows a detail of the LENS time-series for one atom (ID: 144) that slides on the surface along the simulation. Fig~\ref{fig:fig3}E shows the atom's positions at three distinct time frames along with its preceding trajectory, color-coded according to the visited LENS clusters. 
Until $t\sim 40$~ns, the atom remains nearly stationary on the surface (classified in ENV1-2). For $t\gtrsim 40$~ns the atom starts sliding along one of the surface facets (and is classified in ENV3-4: orange, cyan). From $t\sim 125$~ns, the atom is reincorporated into the surface lattice, returning to ENV1-2. Supplementary Movie S3 shows the complete MD trajectory colored based on the detected clusters. 

Fig~\ref{fig:fig3}F shows how 6/5 LENS clusters can be clearly classified with a negligible information loss up to $\Delta t \sim 0.1-0.2$~ns time resolution. However, such atomic sliding events are so rapid that for larger $\Delta t$ these get lost in the analysis. From $\Delta t > 0.3$~ns the total number of LENS clusters diminishes to 2, and the algorithm can distinguish only the static (ENV1) from the non-static (ENV0) atoms. 

\subsubsection*{Analysis of multivariate time-series}
While the examples above show the efficiency of {\em Onion Clustering} in analysing univariate (uni-dimensional) time-series, in many case it is desirable to conduct multidimensional analyses to minimize information loss. We thus extended the method to made it capable of processing also multivariate time-series. The main adaptation concerns the use of a multivariate Gaussian distribution for fitting the histogram maxima. As a proof of efficiency, we thus tested the method on prototypical examples of $2-$ or $3-$dimensional time-series data, using a factorized Gaussian distribution (see Methods section for details). 

\begin{figure*}[!ht]
    \centering
    \includegraphics[width=\textwidth]{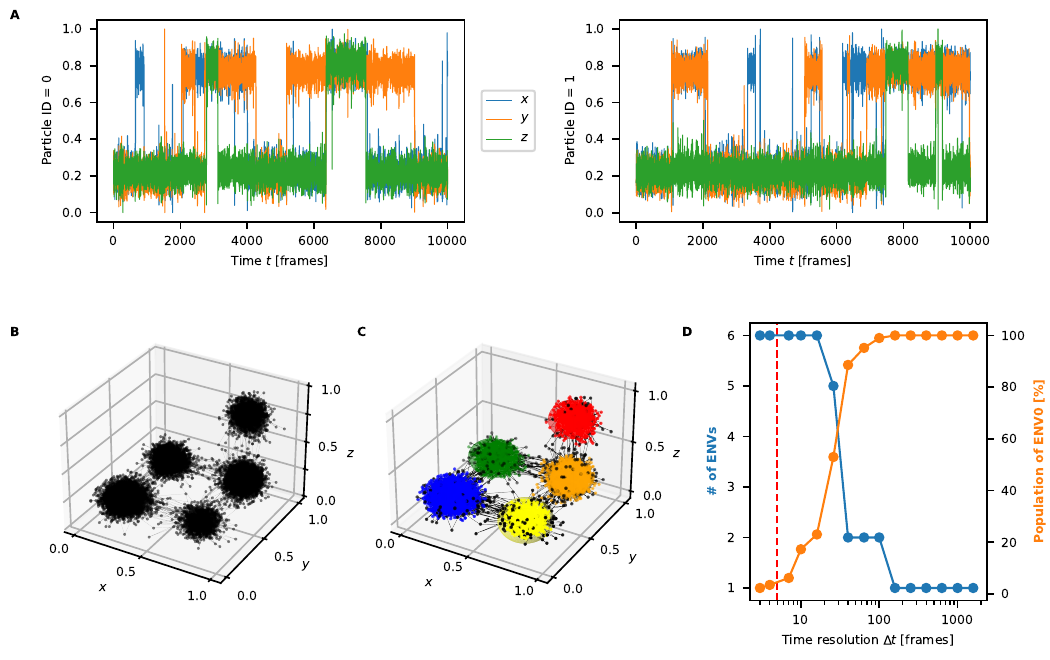}
    \caption{\textbf{Analysis of synthetic multivariate time-series.} A: Time-correlated data-points where generated simulating two particles (shown on the left and right panel) with Langevin Dynamics in a three-dimensional free-energy landscape with 5 minima. B: the same trajectories, represented as a three-dimensional signal. C: The output of the clustering algorithm. The identified clusters are represented as ellipsoidal surfaces, including the points closer than $2\sigma$ from the center. D: Blue line: number of clusters identified as a function of the time resolution $\Delta t$. Orange line: percentage of the data points classified in the ENV0 cluster as a function of the time resolution $\Delta t$. The red dashed line at $\Delta t = 5$~frames indicates at which time resolution the analysis shown in the previous panels was performed. }
    \label{fig:fig4}
\end{figure*}

As a first test case, we constructed a synthetic $3-$dimensional time-series data, generated by simulating $N=2$ non-interacting particles. The particles move {\it via} Langevin dynamics in a three-dimensional free energy landscape featuring 5 distinct minima. Shown in Fig~\ref{fig:fig4}A-B, such a simple dataset shows $5$ clear maximum density clusters separated by sparse data points. As illustrated in Fig~\ref{fig:fig4}C, {\em Onion Clustering} effectively detects the 5 clusters (results obtained using $\Delta t = 5$~frames). 

The plot of Fig~\ref{fig:fig4}D shows the impact of varying the time resolution $\Delta t$. The correct number of clusters is accurately identified up to $\Delta t = 16$~frames. Reducing the resolution and using a larger $\Delta t$ in the analysis of the time-series, the clusters start to merge leading to a clear information loss. In fact, the population of ENV0 remains $< 10\%$
up to $\Delta t = 7$~frames, while beyond this limit it increases rapidly. 

This simple example shows how {\em Onion Clustering} can be used also to analyze in general multivariate time-series data. This includes also dataset that are less artificial and more noisy than this synthetic example, and not necessarily coming from simulated trajectories, as discussed in the next section. 

\subsubsection*{{\em Onion Clustering} of experimental multidimensional time-series}
\begin{figure*}[!ht]
    \centering
    \includegraphics[width=\textwidth]{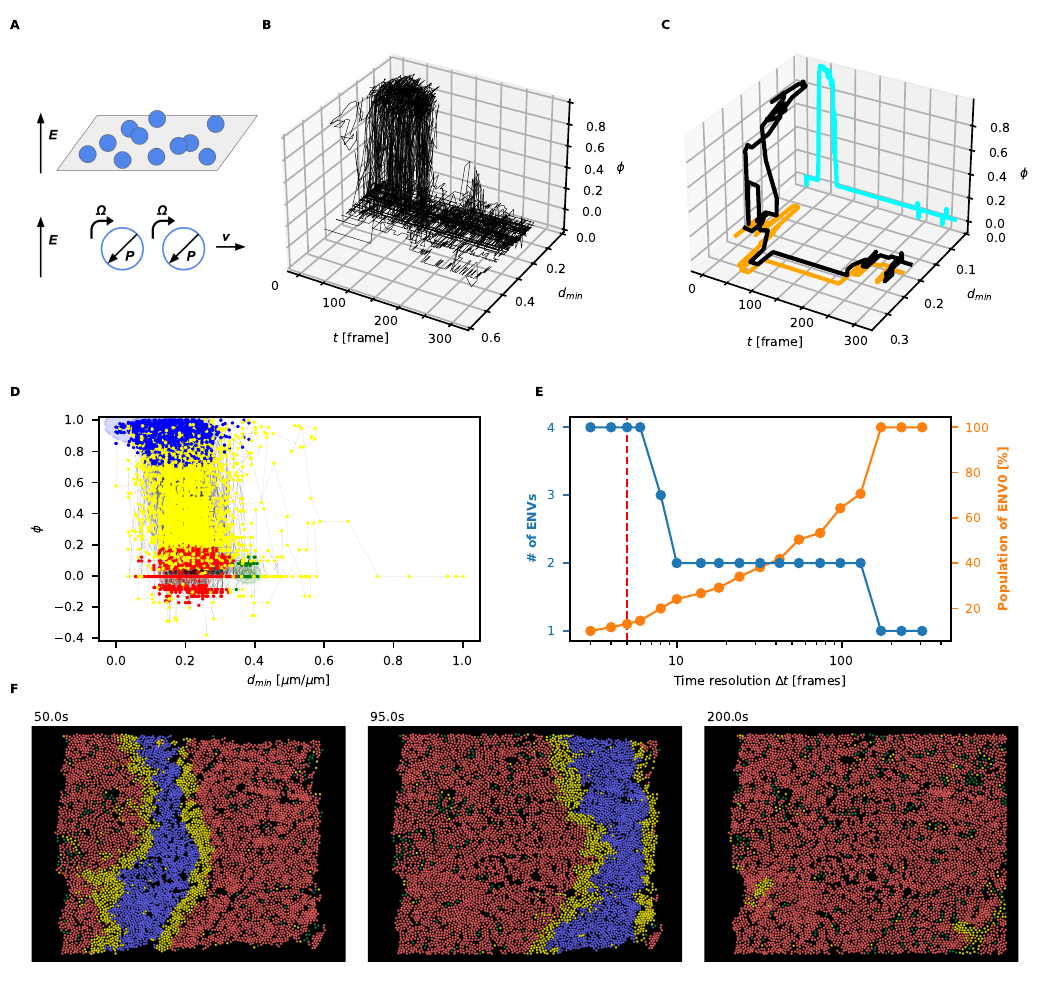}
    \caption{\textbf{Analysis of experimental multivariate time-series}. A. Cartoon representation of Quincke rollers, dielectric colloidal particles suspended in a conducting fluid and exposed to a vertical DC electric field. These particles exhibit collective motion, see Liu {\it et al}~\cite{liu2021activity} for more information. B: The rescaled minimum neighbor distance $d_{min}$ and the local velocity alignment $\phi$ are plotted as a function of time, for all the particle in the video. C: Example signal for a single particle is shown (in black), together with its two separate components (in orange and cyan). D: The algorithm identifies three clusters, in red, blue and green respectively. E: Blue line: number of clusters identified as a function of the time resolution $\Delta t$; orange line: percentage of the data points classified in the ENV0 cluster as a function of the time resolution $\Delta t$. The red dashed line at $\Delta t = 5$~frames indicates the time resolution for the show results. F: Three snapshots from the video, colored according the detected clusters. }
    \label{fig:fig5}
\end{figure*}

As a last example, we tested {\em Onion Clustering} onto multivariate experimental time-series data sourced from a recent study of the complex dynamics of colloidal Quincke roller particles by Liu {\it et al}~\cite{liu2021activity}. Briefly, Quincke rollers are $\mu$m scale dielectric colloidal particles suspended in a conducting fluid and exposed to a vertical DC electric field (see Fig~\ref{fig:fig5}A). 
While for a detailed description of these systems we refer the reader to the original publication, what is interesting to us here is that, under the stimulus of the electric field, these particles exhibit complex collective motions, eventually manifesting as collective density waves or vortexes. Noteworthy, unlike the molecular-scale examples, this test deals with a complex mesoscopic system, and the data originate from experimental observations rather than from simulations. 
As a proof of concept, we considered an optical microscope movie tracking $N = 6921$ particles in a field of view is 700 × 700 $\mu m^2$ for 0.25 s of real time (for a total of $T = 310$~frames), where a collective density wave emerges and runs in the system~\cite{liu2021activity}.  From this movie we extracted the particles' positions at each time-frame along the trajectory using the python package Trackpy~\cite{trackpy, crocker1996methods}. 

For each particle in the system, we extracted from the trajectory data at each sampled frame (i) the minimum neighbor distance ($d_{\text{min}}$: a proxy for the local particle density), and (ii) the particles' local velocity alignment, computed as:
\begin{equation}
    \phi_i \equiv \frac{1}{n_c^i} \sum_j \frac{\Vec{v}_i\cdot\Vec{v}_j}{|\Vec{v}_i| |\Vec{v}_j|}
    \label{eq:eq3}
\end{equation}
In Eq~\ref{eq:eq3}, $j$ iterates over the $n_c^i$ particles inside a specified cutoff distance $r_c$ from particle $i$ ($r_c = 15$~pixels). $\Vec{v}_i$ and $\Vec{v}_i$ are the velocities of particles $i$ and $j$, respectively. The variable $\phi_i$ captures the average cosine value of the angle between the velocities of particle $i$ and its neighboring particles: this value ranges between $-1$ and $1$, indicating the level of alignment or orientation similarity between the velocities of the particle and its neighbors. 

We thus obtained the bi-dimensional time-series data shown in Fig~\ref{fig:fig5}B (showing the time-series for all particles) and Fig~\ref{fig:fig5}C (showing a single particle, in black, and its two components (i) and (ii), in orange and cyan) over time. $d_{\text{min}}$ was rescaled within the range $[0,1]$ to facilitate the visualization and give the to variables (i) and (ii) a comparable weight. 

Fig~\ref{fig:fig5}D-F show as an example the results obtained by {\em Onion Clustering} employing a time resolution of $\Delta t = 5$~frames. The algorithm identifies 3 distinct statistically-relevant environments (blue, red and green) alongside the ENV0 cluster encompassing all unclassified data points (in yellow).
The significance of the clusters becomes apparent when observing the simulation snapshots in Fig~\ref{fig:fig5}F and Supplementary Movie~S4. Environment ENV1 (in red) is characterized by $d_{\text{min}} = (0.20\pm0.08)$ and $\phi = (0.01\pm0.10)$, and primarily consists of stationary particles. ENV2 (blue) is characterized by $d_{\text{min}} = (0.12\pm0.15)$ and $\phi = (0.97\pm0.14)$, and corresponds to particles moving coherently within the wavefront. ENV3 (green) contains particles with $d_{\text{min}} = (0.38\pm0.04)$ and $\phi = (0.01\pm0.06)$, stationary particles located in an area with very low density (exhibiting high $d_{\text{min}}$ values). The unclassified data points (ENV0: yellow) correspond to particles situated on the two edges of the wave, whose surrounding environment is changing too rapidly to be classified as persistent clusters at this time-resolution. 

Reducing $\Delta t$ reduces the population of the ENV0 cluster and increases the ability of the algorithm to precisely characterize the edges of the wave (see Fig~S4). Fig~\ref{fig:fig5}E shows the number of clusters and the population of ENV0 as a function of the $\Delta t$. Notably, 4 clusters are discernible when employing $\Delta t \leq 6$ frames, a timescale comparable to the residence time of a single particle inside the wave. 

\section*{Conclusions}
In this paper we introduced {\em Onion Clustering}, a new unsupervised clustering algorithm for the microscopic analysis of time-series data. {\em Onion Clustering} automatically identifies fluctuations and microscopic dynamic environments in a time-series, and classifies the data points into micro-clusters. Typically, unsupervised clustering methods suffer, {\it e.g.}, of lack of physical interpretability of the results, multiple parameters that have to be tuned (and that may considerably change the results), and difficulties in identifying clusters/environments that are way less sampled/populated than others, such as rare and/or local dynamical events, transient states, etc. Here, using various types of test examples, we demonstrate how {\em Onion Clustering} can mitigate such issues, standing out as a general and reliable unsupervised method characterized by non-common physical interpretability, statistical robustness, ease of use, and flexibility in analyzing different types of time-series data. 

{\em Onion Clustering} is based on an iterative ``certainty-based'' approach. The most evident and statistically populated environment is classified first, and then it is removed, together with its noise, from the time-series, which is then analyzed again in iterative fashion. The algorithm can thus rely on an adaptive metric that, at every successive iteration, enhances the signal-to-noise ratio. In an ``onion peeling'' fashion, this allows to unveil all the dynamical subdomains (also the least populated ones) that can be classified in a statistically-robust way at a given time-resolution. In this way, the method can extract and retain all information information that are statistically significant in a time-series as a function of the resolution at which this is studied. 
At the same time, {\em Onion Clustering} quantifies -- {\it via} the population of the ENV0 cluster ({\it i.e.}, the data points which was not possible to classify) -- the amount of information that cannot be statistically classified and that gets lost at a certain time resolution $\Delta t$, which is a non-trivial added value for an unsupervised method. 
While in such a method, the time-resolution $\Delta t$ is the sole determinant parameter, instead of choosing the time-resolution {\it a priori}, {\em Onion Clustering} performs the analysis at every possible resolution (the bottom limit being the time-interval between the frames in the time-series itself) and plots the results. This allows the user to make an {\it a posteriori} informed choice of the resolution at which it is best to study a time-series to analyze determined phenomena/events. This makes {\em Onion Clustering} a fully unsupervised, substantially parameter-free clustering method that is transparent, controllable, statistically robust, and that avoids typical problems emerging from the use of such unsupervised algorithms as a black box. 

The examples discussed herein demonstrate how {\em Onion Clustering} can efficiently reconstruct all the statistically-relevant events contained in time-series with extremely variegated features: from systems in dynamical equilibrium conditions, to systems out-of-equilibrium, to systems dominated by rare events and local fluctuations (difficult to detect by pattern recognition analyses due to their negligible statistical weight), from synthetic and simulation time-series, to experimental trajectories. 
We expect that, thanks to its generality and simplicity, {\em Onion Clustering} will constitute a useful tool in the study of complex systems from the atomistic to the macroscopic scale. 

\section*{Methods}
\subsection*{Simulations and data analysis}
\subsubsection*{Water-ice in dynamic coexistence}
The data for Fig~\ref{fig:fig1} (available in the SI: Dataset S1) are obtained by the MD simulations described in details in~\cite{crippa2023detecting, caruso2023timesoap, crippa2023machine}, of 2048 TIP4P/ICE molecules in correspondence of the melting temperature ($T=268$~K for this force field~\cite{conde2017high}). In the initial configuration, half of the molecules are in the solid phase in hexagonal ice packing, the other half are in the liquid phase. Being at the melting temperature, solid and liquid phase are in dynamical equilibrium. The simulation lasts $50$~ns with a configuration sampling interval of $0.1$~ns. For every molecule in the system, the LENS signals are computed on the sampled configurations using a cutoff of $10$~\AA~(close to the third minimum of the radial distribution function), for the 2048 oxygen atoms~\cite{crippa2023detecting, crippa2023machine}. 

\subsubsection*{Water freezing}
The data for Fig~\ref{fig:fig2} (available in the SI: Dataset S2) are obtained by continuing the simulation at coexistence using the same setup, but lowering the temperature to $T=267$~K (just below the TIP4P/ICE melting point~\cite{conde2017high}). In this case, the system is evolved for $40$~ns, sampling its configuration every ps, while only the last $5$~ns of the simulation were used in the analysis: right before the freezing started, in such a way to have a prevalence of signal related to the ice domain and in order to test the algorithm in an unfavorable case, and to prove that this can keep track that some liquid water has been present in the trajectory even in the case where this has a low statistical weight in the time-series. The $t$SOAP signals~\cite{caruso2023timesoap} are computed, with a cutoff of $10$~\AA~(close to the third minimum of the radial distribution function), for the 2048 oxygen atoms. The $t$SOAP signals are then smoothed with a moving average with $25$~frames width (to reduce the noise). 

\subsubsection*{FCC(211) copper surface}
The data for Fig~\ref{fig:fig3} (available in the SI: Dataset S3) are obtained by the MD simulation described in details in~\cite{cioni2023innate}, of 2400 Cu atoms at $T=600$~K . Periodic boundary conditions are applied in the $xy$ plane, while the system is finite along the $z$ direction, simulating {\it de facto} an infinte copper/vacuum surface along the (211)-plane. The system is evolved for $150$~ns, sampling its configuration every $10$~ps, for a total of 15000~frames. 
The LENS signals~\cite{crippa2023detecting} are then computed, with a cutoff of $6$~\AA, for all the atoms. The signals are then smoothed with a moving average with $10$~frames width (to reduce the noise). 

\subsubsection*{Multivariate/multidimensional synthetic data}
The data for Fig~\ref{fig:fig4} (available in the SI: Dataset S4) are obtained simulating 2 non-interacting particles with Langevin dynamics, in a bounded potential energy surface with five minima, with coordinates $(0,0,0)$, $(0,1,0)$, $(1,0,0)$, $(1,1,0)$ and $(1,1,0)$. The system is evolved for $2\cdot10^6$~integration steps, sampling its configuration every $200$~steps. Particles' coordinates are then given as input to the clustering algorithm. 

\subsubsection*{Experimental trajectories of Quincke rollers}
The data for Fig~\ref{fig:fig5} (available in the SI: Dataset S5) are obtained {\it via} image recognition and a tracking code (trackpy~\cite{trackpy}) from experimental microscopy videos from~\cite{liu2021activity}. From the video, the $x-$ and $y-$coordinates of 6921 particles for 310 consecutive frames are extracted. For each particle at each frame, the distance from the closest neighbor $d_{min}$ and the local alignment of the velocities $\phi$ (as defined in the main text, with a cutoff distance of $r_c = 15$~pixels) are computed. The two quantities are then separately smoothed with a moving average with $2$~frames width (to reduce the noise). 

\subsection*{The clustering algorithm}
\subsubsection*{Univariate/monodimensional data analysis}
Let's call $x_i(t)$, with $0\leq i < N$ indexing the particle and $0\leq t < T$ indexing the discrete time, the set of signals we want to cluster. The algorithm proceeds as follow:
\begin{enumerate}
    \item The signals $x_i(t)$ are divided in windows of length $\Delta t$, the time resolution of the analysis:
    \begin{equation*}
        \resizebox{0.43\textwidth}{!}{$X_{i, w} = \left[ x_i(w\Delta t), x_i(w\Delta t + 1), x_i(w\Delta t + 2), ..., x_i(w\Delta t + \Delta t - 1)\right]$}
    \end{equation*}
    with $w\in \{0, 1, 2, ...,  \mbox{int}(T/\Delta t)\}$. 
\end{enumerate}
The following procedure is then repeated iteratively, each time identifying a candidate environment $E_n$, until a termination condition is met:
\begin{enumerate}
    \setcounter{enumi}{1}
    \item The cumulative histogram $H_j$ of all the data is computed, with $0\leq j < n_{bins}$. $n_{bins}$ is set automatically~\cite{numpy_binning}, but can be also set to a custom value. 
    \item The absolute maximum of the histogram is identified, and a Gaussian distribution of the form Eq~\ref{eq:eq1} is fitted on the histogram in an interval around the maximum, with $\mu$, $\sigma$ and $A$ as free parameters. The details about the choice of the fitting interval are reported in SI. If the fitting procedure does not converge, go to step 7. 
    \item A candidate environment $E_n$ is identified as the signal interval 
    $$E_n = [\mu_n - 2\sigma_n, \mu_n + 2\sigma_n]$$
    The values of $\mu_n, \sigma_n$ and $A_n$ are stored for later use. 
    \item For every pair $(i, w)$, the window $X_{i, w}$ is removed from the signals if and only if it's entirely included in the environment $E_n$, that is, if and only if
    \begin{equation}
        \left\{\begin{array}{@{}l@{}}
            \min{\{X_{i, w}\}} \geq \mu_n - 2\sigma_n\\
            \max{\{X_{i, w}\}} \leq \mu_n + 2\sigma_n
        \end{array}\right.\,
    \end{equation}
    If no window satisfies this requirements, go to step 7. 
    \item If after this step the signals $x_i(t)$ are still not empty, the procedure is repeated from the step 2. Otherwise, go on to step 7. 
    \item At this point, a list of environment $E_n$, $0\leq n < n_{\mbox{states}}$, has been identified, each one described by its center $\mu_n$, its width $4\sigma_n$ and its weight $A_n$. Moreover, a fraction $f_n$ of windows $X_{i, w}$ has been assigned to each environment. From this information, strongly overlapping environments are merged together. The details about this procedure are reported in SI. After this, each data point $x_i(t)$ is assigned to the environment $i$ in which its window is contained. 
\end{enumerate}

\subsubsection*{Multivariate/multidimensional data analysis}
\label{met:part2}
The case of $D-$dimensional signals is handled in basically the same way as in one-dimensional ones. The Gaussian used of the fit around the maxima are factorized, {\it i. e.} of the form
$$P(x_1, x_2, ..., x_D) = \prod_{d=1}^D\frac{A_d}{\sqrt{\pi}\sigma_d}\exp{\left[-\left(\frac{x_d - \mu_d}{\sigma_d}\right)^2\right]}$$
and the fit is performed inside a $D-$dimensional rectangular region, where the limit of the rectangle along each dimension are selected with the same procedure shown in SI for the univariate data.

\section*{Acknowledgements}
G.M.P. acknowledges the funding received by the European Research Council (ERC) under the European Union’s Horizon 2020 research and innovation program (grant agreement no. 818776--DYNAPOL). 

\section*{Data availability}
The algorithm presented in this paper is implemented as a Python3 package~\cite{onion-pypi}. The code is available open-source at this GitHub repository~\cite{onion-git, gmp-git}. All the relevant code, Supporting Datasets and Movies for this work are available on Zenodo at~\cite{zenodo-onion}. 

\section*{Competing interests statement}
The authors declare no competing interests.

\printbibliography

\newpage
\section*{Supporting Information for:}

{\Large ``Layer-by-layer'' unsupervised clustering of statistically relevant fluctuations in noisy time-series data of complex dynamical systems}

\vspace{0.5cm}
\noindent Matteo Becchi\footnotemark[1]{}, Federico Fantolino\footnotemark[1], Giovanni M. Pavan\footnotemark[1]\footnotemark[2]

\footnotetext[1]{Department of Applied Science and Technology, Politecnico di Torino, Torino 10129, Italy}
\footnotetext[2]{Department of Innovative Technologies, University of Applied Sciences and Arts of Southern Switzerland, Lugano-Viganello 6962, Switzerland}

\subsection*{Further algorithm details}

\subsubsection*{Identification and selection of the fitting interval}
\label{sec:appendix_A}
The identification of the correct fitting interval surrounding the histogram maximum is often critical to ensure not only the convergence, but also the quality of the Gaussian fit. Our algorithm selects the best fitting interval between two candidate intervals defined as follow:
\begin{itemize}
    \item The ``\textit{minima}'' interval is defined by the positions of the two local minima immediately before and after the maxima. 
    \item The ``\textit{half height}'' interval is determined by identifying the points on either side of the maximum in the histogram where the histogram reaches half of its maximum height. If the edge of the histogram is reached before finding a point with half height, the interval will extend up to the edge. 
\end{itemize}
The fit is then attempted inside both candidate intervals, and (if both fits converge) a quality score is assigned to the intervals, according to the following desired properties:
\begin{itemize}
    \item the value of the maximum of the fitting Gaussian is close enough to the value of the maximum of the histogram;
    \item the Gaussian mean $\mu$ is inside the fitting interval;
    \item the Gaussian width $\sigma$ is smaller than the fitting interval;
    \item the relative uncertainty on each fitting parameter is smaller than 0.5. 
\end{itemize}
Finally, the fitting interval with the highest quality score is selected. If only one of the fits converges, that one is selected. If none of them converges, a termination condition is met and the algorithm exits the iterative loop.

\subsubsection*{Removing strongly overlapping clusters}
\label{sec:appendix_B}
Once the algorithm has identified the set of candidate environments $\{E_n\}$, all the possible pair of environments $E_n$ and $E_m$ are compared. If 

$$\frac{A_n}{\sigma_n} > \frac{A_m}{\sigma_m}$$
(meaning, $E_n$ has a peak higher than $E_m$), and

$$|\mu_n - \mu_m| < \sigma_n$$
(meaning, $E_m$ is closer to $E_n$ then $E_n$'s typical fluctuation amplitude), then $E_m$ is considered contained in $E_n$. $E_m$ is thus removed from the list of candidate environments, and all the data points previously classified inside $E_m$ are now consider classified inside $E_n$. If $E_m$ meets the criteria for being contained in more that one different environment, the one with the mean closer to $\mu_m$ is chosen. 

\setcounter{figure}{0}
\renewcommand{\figurename}{Fig.}
\renewcommand{\thefigure}{S\arabic{figure}}

\begin{figure}
    \centering
    \includegraphics[width=\textwidth]{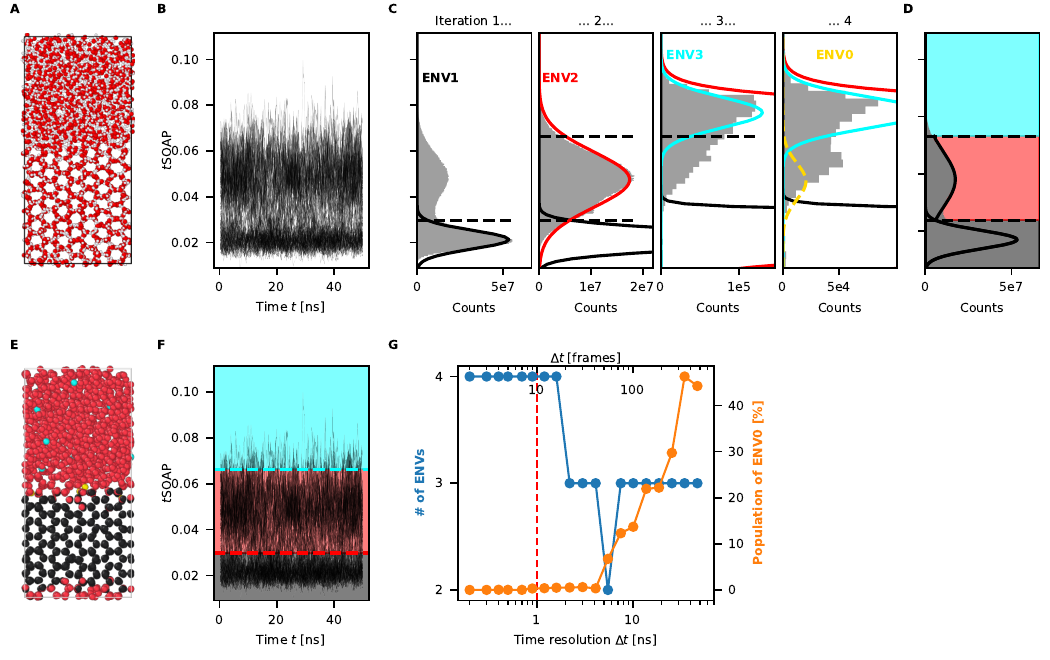}
    \caption{\textbf{Clustering on $t$SOAP signals with $\Delta t = 10$~frames.} A: Snapshot of the simulation of ice/water coexistence; the simulation is performed on 2048 TIP4P/ICE molecules, lasts $50$~ns and it's sampled every $0.1$~ns. B: $t$SOAP signals for all the oxygen atoms, as a function of time. Data are smoothed with a rolling average with width of 8 frames. C: Data cumulative distribution function at the four iterations of the algorithm. The solid lines are the Gaussian best fit of the maximum of the distribution. The dashed lines are the threshold between the identified clusters. At the fourth iteration, no data are assigned to the proposed cluster, and the algorithm stops. D: final clustering of the $t$SOAP signals. E: Snapshot of the simulation, colored according to the clustering. F: Same $t$SOAP signals of panel B; background is colored according to the thresholds given by the clustering algorithm. G: Blue line: number of clusters identified as a function of the time resolution $\Delta t$; orange line: percentage of the data points classified in the ENV0 cluster as a function of the time resolution $\Delta t$. The red dashed line at $\Delta t = 1$~ns (10 frames) indicates at which time resolution the analysis shown in the previous panels was performed. }
\end{figure}

\begin{figure}
    \centering
    \includegraphics[width=\textwidth]{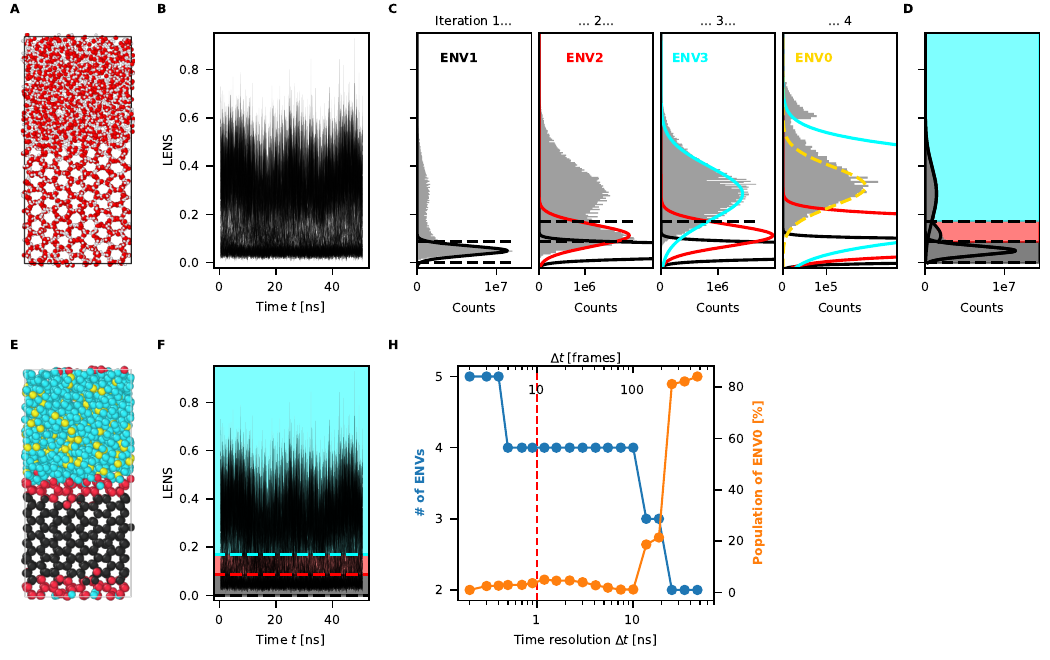}
    \caption{\textbf{Clustering on LENS with $\Delta t = 10$~frames.} A: Snapshot of the simulation of ice/water coexistence; the simulation is performed on 2048 TIP4P/ICE molecules, lasts $50$~ns and it's sampled every $0.1$~ns. B: LENS signals for all the oxygen atoms, as a function of time. C: Data cumulative distribution function at the four iterations of the algorithm. The solid lines are the Gaussian best fit of the maximum of the distribution. The dashed lines are the threshold between the identified clusters. At the fourth iteration, no data are assigned to the proposed cluster, and the algorithm stops. D: final clustering of the LENS signals. E: Snapshot of the simulation, colored according to the clustering. F: Same LENS signals of panel B; background is colored according to the thresholds given by the clustering algorithm. G: Blue line: number of clusters identified as a function of the time resolution $\Delta t$; orange line: percentage of the data points classified in the ENV0 cluster as a function of the time resolution $\Delta t$. The red dashed line at $\Delta t = 1$~ns (10 frames) indicates at which time resolution the analysis shown in the previous panels was performed. }
\end{figure}

\begin{figure}
    \centering
    \includegraphics[width=\textwidth]{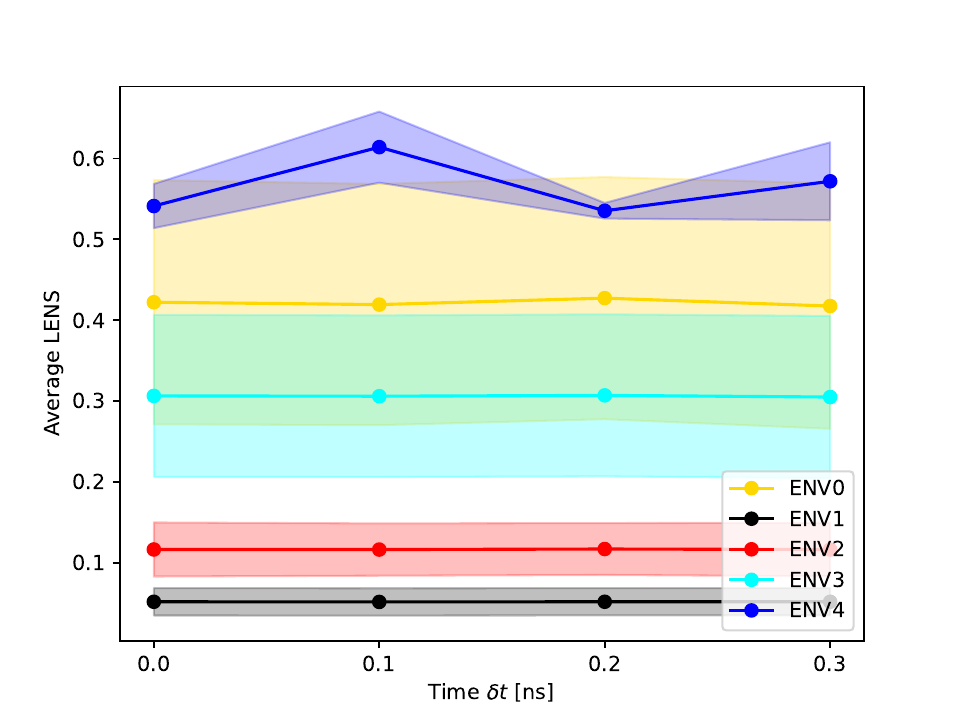}
    \caption{\textbf{Characterization of the clusters shown in Fig1 in the main text.} The plot shows, for each cluster (ENVs0-4), the average, over all the windows classified in that cluster, of the LENS signal in the consecutive frames inside the window. The data refers to the clustering performed with a time-resolution $\Delta t = 0.4$~ns ({\it i.e.}, 4 simulation frames). The solid dots are the average, the transparent bands the standard deviation. It can be seen that the clusters ENV1 (in black, corresponding to solid ice), ENV2 (in red, corresponding to solid/liquid interface), ENV3 (cyan, corresponding to the majority of the liquid water) and ENV4 (blue, corresponding to liquid water with high values of LENS) are well separated with respect to their standard deviation. The ENV0 cluster instead (in yellow) includes data windows with high variability, which reflects in its large standard deviation. }
\end{figure}

\begin{figure}
    \centering
    \includegraphics[width=\textwidth]{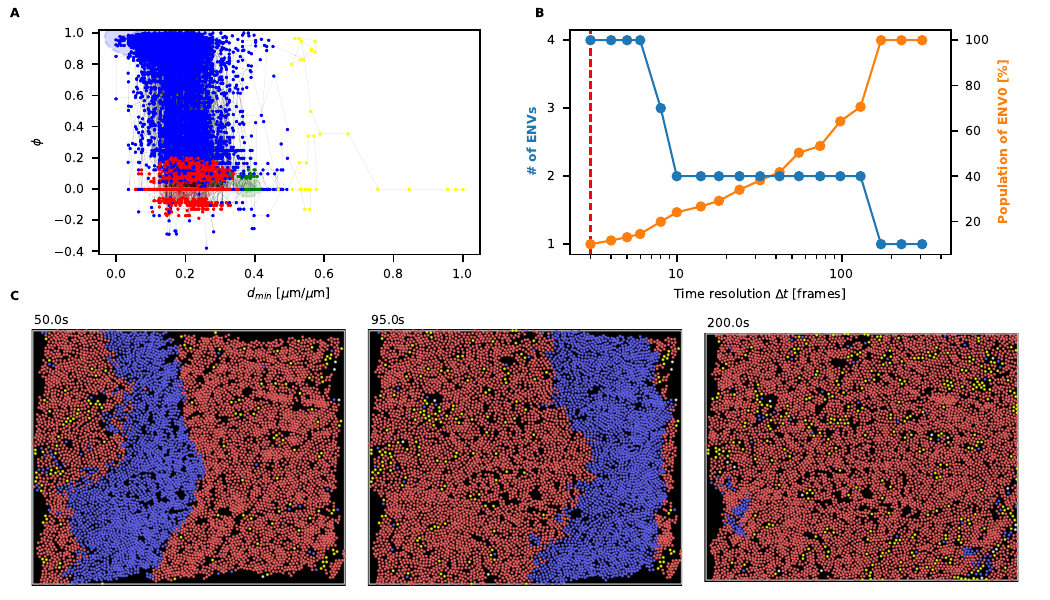}
    \caption{\textbf{Analysis of multivariate data from experimental microscopy video of colloidal particles, using $\Delta t = 3$~frames. Compare with Fig5 in the main text.} A: the algorithm identifies three environments, in red, blue and green respectively, plus the ENV0 cluster in yellow. E: Blue line: number of clusters identified as a function of the time resolution $\Delta t$; orange line: percentage of the data points classified in the ENV0 cluster as a function of the time resolution $\Delta t$. The red dashed line at $\Delta t = 3$~frames indicates at which time resolution the analysis shown in the previous panels was performed. F: three snapshots of the video, colored according the clustering output.}
\end{figure}

\newpage
\paragraph{Movie S1} MD trajectory of solid/liquid water coexistence used for the analysis of Fig1 in the main text, colored according to the clustering with $\Delta t = 0.4$~ns. 5 clusters are found; the black one (ENV1) correspond to solid ice, the red one (ENV2) mainly to the solid/liquid interface, the cyan (ENV3) and blue (ENV4) ones to liquid water, and the yellow one (ENV0) to the unclassified points. 

\paragraph{Movie S2} The MD trajectory of freezing water at $T = 267$~K used for the analysis of Fig2 in the main text, colored according to the clustering with $\Delta t = 25$~ps. 5 clusters are found; the black one (ENV1) correspond to solid ice, the red one (ENV2) mainly to the solid/liquid interface, the cyan (ENV3) and blue (ENV4) ones to liquid water, and the yellow one (ENV0) to the unclassified points. 

\paragraph{Movie S3} The MD trajectory of copper 211 surface used for the analysis of Fig3 in the main text, colored according to the clustering with $\Delta t = 0.12$~ns. 5 clusters are found; the black one (ENV1) correspond to the atoms in the bulk, the red one (ENV2) to a fraction of the surface atoms, the orange (ENV3) and cyan (ENV4) ones to atoms sliding on the surface, and the yellow one (ENV0) to the unclassified points. 

\paragraph{Movie S4} Reconstruction of the experimental video used for the analysis of Fig5 in the main text, colored according to the clustering with $\Delta t = 5$~frames. 4 clusters are found; the red one (ENV1) correspond to the majority of stationary rollers, the blue one (ENV2) to rollers inside the wave moving in a coherent and ordered way, the green (ENV3) one to stationary rollers in areas with exceptionally low particle density, and the white one (ENV0) to the unclassified points, mainly located before and after the wave. 

\paragraph{Dataset S1: timeseries\textunderscore Fig1.npy} LENS signals for Fig1 of the main text. The file contains an array of shape $(N, T)$, where $N = 2048$ is the number of TIP4P/ICE molecules and $T=500$ is the number of simulation frames. 

\paragraph{Dataset S2: timeseries\textunderscore Fig2.npy} $t$SOAP signals for Fig2 of the main text. The file contains an array of shape $(N, T)$, where $N = 2048$ is the number of TIP4P/ICE molecules and $T=40000$ is the number of simulation frames. 

\paragraph{Dataset S3: timeseries\textunderscore Fig3.npz} LENS signals for Fig3 of the main text. The file contains an array of shape $(N, T)$, where $N = 2400$ is the number of Cu atoms and $T=15270$ is the number of simulation frames. 

\paragraph{Dataset S4: timeseries\textunderscore Fig4.npy} Synthetic signals for Fig4 of the main text. The file contains an array of shape $(D, N, T)$, where $D = 3$ is the number of signal components, $N = 2$ is the number of Langevin molecules and $T=10000$ is the number of simulation frames. 

\paragraph{Dataset S5: timeseries\textunderscore Fig5.npy} $d_\text{min}$ and $\phi$ signals for Fig5 of the main text. The file contains an array of shape $(D, N, T)$, where $D = 2$ is the number of signal components, $N = 6921$ is the number of colloidal particles and $T=311$ is the number of video frames. 

\end{document}